# Optimal Indexes for Sparse Bit Vectors*

Alexander Golynski[1], Alessio Orlandi[2], Rajeev Raman[3], and S. Srinivasa Rao[4]

[1] Google Inc., New York City, U.S.
agolynski@google.com
[2] Google Switzerland GmbH, Zürich, Switzerland
oalessio@google.com
[3] Department of Computer Science, University of Leicester, U.K.
r.raman@mcs.le.ac.uk
[4] School of Computer Science and Engineering, Seoul National University, S. Korea.
ssrao@cse.snu.ac.kr

**Abstract.** We consider the problem of supporting rank and select operations on a bit vector of length $m$ with $n$ 1 bits. The problem is considered in the *succinct index* model, where the bit vector is stored in "read-only" memory and an additional data structure, called the *index* is created during pre-processing to help answer the above queries. We give asymptotically optimal density-sensitive trade-offs, involving both $m$ and $n$, that relate the size of the index to the number of accesses to the bit vector (and processing time) needed to answer the above queries. The results are particularly interesting for the case where $n = o(m)$.

## 1 Introduction

We consider the problem of representing a bit vector $S[1..m]$ of length $m$, and supporting the following operations, for $x \in \{0, 1\}$:

– $\mathsf{rank}_x(S, i)$ returns the number of occurrences of $x$ in the prefix $S[1..i]$.
– $\mathsf{select}_x(S, i)$ returns the position of the $i$th occurrence of $x$ in $S$.

Such a data structure is called a *fully indexable dictionary (FID)* [25]. We consider this problem in the context of *systematic* encodings, also known as the *succinct index* model. In this model, the bit vector $S$ is not directly accessible to the data structure as "bits in memory": instead, when answering a query, the data structure can inspect parts of $S$ through an access operation, which may be relatively expensive. In order to reduce the number of access operations, we augment $S$ with an *index*, or a data structure that contains pre-computed information specific to $S$. To answer a query, the data structure performs a combination of access operations and "local" computations using the index as well as results of the access operations. We consider three measures of the performance of the data structure:

---

* Some of these results were published in preliminary form in the proceedings of the *11th Scandinavian Workshop on Algorithm Theory (SWAT 2008)*, LNCS 5124, Springer. Orlandi's work was done while affiliated with the University of Pisa.

i. The size of the index, also termed the *redundancy*,
ii. the number of access operations,
iii. the amount of "local" computation.

(We do not include preprocessing time, or access operations performed during preprocessing.) We provide new and tight *density-sensitive* lower and upper bounds for this problem, where the above parameters depend both upon the length $m$ and the *weight* $n$ of the bit vector, where the weight is the number of 1s in $S$. One can, without loss of generality, assume that $n \leq m/2$; our primary interest is bit vectors of relatively small weight, i.e., the case $n = o(m)$.

We also mention the *non-systematic* model, as our succinct indices will in fact consist of a number of non-systematic FIDs. In this model, the string $S$ is explicitly given as input. The algorithm is responsible for storing $S$ in an information-theoretic minimum amount of space, according to one of a number of possible measures of the "information content" of $S$. In this setting, the redundancy is the space usage of the data structure above and beyond the information-theoretic minimum amount needed to store $S$.

## 1.1 Motivation

Representing a bit vector to support $\text{rank}_x$ and $\text{select}_x$ is one of the most fundamental operations in the field of space-efficient data structures [24]. Solutions for this problem (such as FIDs) are used in text indexing [15, 16, 6] and representing semi-structured data [9, 5], for example. Considering systematic data structures for this problem appears to have been initiated by the desire to prove lower bounds on index size [8]. However, as the field developed, both algorithmic advantages and technological motivations. Service based computing [?] and remote data storage, two of the main ingredients of nowadays is often referred to as "the cloud", have made the case for systematic data structures stronger. In papers such as [7, 26, 14], the first steps towards recognizing the algorithmic advantages of this approach were taken, and the *succinct index* model was fully formalized (from an upper bound perspective) in [1]. By decoupling the representation of $S$ from the set of operations that are being supported, it is noted in [1] that a succinct index offers many advantages including optimal compressibility of the data being indexed and easy integration of different indices over the same data. Furthermore, a succinct index can also be used in cases when the data being indexed is never explicitly stored, but computed on demand (for example, see [12]).

Finally, the situation that we emphasize here, namely bit vectors of length $m$ with weight $o(m)$, is very important in applications (both practical and algorithmic). Such bit vectors are frequently encountered as the characteristic vectors of sparse sets that naturally occur in practical applications, see e.g. [17]. In addition, many space-efficient data structures often need to represent sequences that have few 1s by construction [15, 9].

### 1.2 Results - Old and New

There have been a number of results on FIDs, and we do not describe all of them here. We note, however, that the redundancy of FIDs has been a focus of research in the non-systematic setting as well, and successive papers [2, 25, 13, 21] have reduced the redundancy of Jacobson's original representation [18]. Very recently, matching non-systematic lower bounds have been shown [23].

Before we describe the systematic/succinct index results for FIDs, we first state the model more precisely. Recall that we want to support operations on a given bit vector $S$ of length $m$, accessible through an access operation. We pre-process $S$ (for free) and create an index $I$ of size $r$ bits. Subsequently:

- For the lower bounds, we assume that access$(i)$, for $1 \le i \le m$, returns the $i$-th *bit* of $S$. When answering the query, the data structure can read $I$ for free, and is only charged for access operations. This model is also referred to as the *bit-probe* model.
- For the upper bounds, we take $\mu = \lceil \lg m \rceil$ assume that access$(i)$, for $1 \le i \le \lceil m/\mu \rceil$, returns the *sequence* of $\mu$ bits of $S$ beginning at position $(i-1)\mu+1$. The "local" computation of the data structure on the index $I$ is analyzed on the RAM model with word size $O(\lg m)$ bits. The performance of the data structure is measured both in terms of the number of access calls it makes as well as the number of operations performed during its "local" computation.

The aim is to study the trade-off between $r$, the size of the index, and the cost of the operations as measured above. In the non-systematic model, upper bounds are also in the word RAM model and only the redundancy and the speed of operations is measured.

We begin by noting that $\mathsf{rank}_0(i) + \mathsf{rank}_1(i) = i$, so a data structure needs to support only one of these operations, and refer to both as rank if this is otherwise immaterial. On the other hand, when we refer to select, it is simply as informal shorthand for referring to "$\mathsf{select}_0$ and/or $\mathsf{select}_1$".

We now summarize existing and new results. A number of recent results give lower bounds on the redundancy of systematic encodings [8, 19, 10]. It has been shown [19, 10] that $\Omega(m \lg \lg m / \lg m)$ redundancy is needed to support FID operations in $O(1)$ time, matched by upper bounds in [10, 25]. Hence, the redundancy of systematic FIDs appears to be a solved question.

The lower and upper bounds, however, are not sensitive to the weight $n$ of the bit vector. For example, when $n = 1$, it is easy to see that redundancy of $O(\lg m)$ bits suffices to support all operations in $O(1)$ time. As already noted, one often has to support FID operations on bit vectors that are constructed to have few 1s. Thus, it is interesting to study the redundancy required to support FID operations as a function of both $m$ and $n$. Previous density-sensitive lower bounds were provided by Golynski [10, Theorem 4.1], who showed that $r = \Omega(n \lg \lg m / \lg m)$ for $t = O(\lg m)$; as we see below, this lower bound is not optimal. Miltersen's [19] work implicitly contains an optimal lower bound for the case $n = \Theta(m/\lg m)$.

| Worst case (over all $n$) | $\Theta(m \lg \lg m / \lg m)$ [19, 10] |
|---|---|
| Density-sensitive [old] | $\Omega(n \lg \lg m / \lg m)$ [10] |
| Density-sensitive [new] | $\begin{cases} \Theta\left(\frac{m}{\lg m} \lg\left(\frac{n \lg m}{m}\right)\right), & \text{if } n = \omega(m/\lg m), \text{ and} \\ \Theta\left(n\left(1 + \lg\left(\frac{m}{n \lg m}\right)\right)\right) & \text{if } n = O(m/\lg m). \end{cases}$ |

**Table 1.** Redundancy of systematic encodings of $O(\lg m)$ bit-probe FIDs.

The lower bound is a complete trade-off that specifies the minimum redundancy required by any data structure that makes at most $t$ bit-probes for any values of $m, n$ and $t$. For simplicity, we focus on the case $t = O(\lg m)$, and the new results are shown in Table 1. As can be seen, the previous lower bound of [10] matches the new ones only for $n = \Theta(m)$ (i.e. for the dense case). For the case $n = \Theta(m/\lg m)$ for instance, the new lower bound is $\Omega(n) = \Omega(m/\lg m)$, whereas the old lower bound was $\Omega(n \lg \lg m / \lg m) = \Omega(m \lg \lg m / (\lg m)^2)$.

We show matching upper bounds, giving succinct indices that support all FID operations, perform $O(1)$ access operations and support queries in $O(1)$ time if, additionally, $m/n = (\lg m)^{O(1)}$. Note that in the upper bound model, a single access operation reads $\lg m$ consecutive bits of $S$, so the lower and upper bounds are indeed comparable (indeed, the lower bound allows the algorithm to probe $O(\lg m)$ *arbitrary* bits per query, not just consecutive ones). Furthermore, the restriction in the upper bound that the "local" computation takes constant time only when $m/n = (\lg m)^{O(1)}$ is due to the lower bound on predecessor queries in the RAM model [22] (recall that the lower bound assumes that all "local" computation is for free).

The rest of the paper is organized as follows. Section 2 contains the upper bound and begins with a (non-systematic) data structure that represents a bit vector of length $m$ with weight $n$ in $B(m, n) + O(\min\{n, m - n\}) = O(B(m, n))$ bits and supports only $\mathsf{select}_0$ and $\mathsf{select}_1$ in $O(1)$ time, where $B(m, n) = \lg \lceil \binom{m}{n} \rceil$ (provided $\min\{n, m - n\} = m/(\lg m)^{O(1)}$). In fact, FIDs are known that can achieve a space bound of $B(m, n)(1 + o(1))$ bits [13, 21], but they are significantly more complex; the experimental work of [20] suggests that our approach is practical. We then use this data structure in our systematic index. The lower bounds are described in Section 3 and are based upon the general *choices tree* framework of [10].

## 2 Upper Bounds

### 2.1 Non-systematic FIDs for moderately-dense bit sequences

We begin by giving a number of results on non-systematic FIDs; these FIDs will be heavily used in our succinct index. In this section We begin by stating a classical result on FIDs, due to Clark and Munro [2]:

**Lemma 1.** *There is a FID that stores a bit vector $S$ of length $m$ using $m + o(m)$ bits and supports all operations in $O(1)$ time.*

Next, we show the following lemma, which is essentially the *Elias-Fano* representation of a sequence [4], straightforwardly augmented with the rank operation in $O(1)$ time (a related data structure was shown to have good practical performance in [20]):

**Lemma 2.** *We can store a bit vector of length $m$ and weight $n$, providing that $n \geq m/(\lg m)^c$ for some constant $c > 0$, such that we can support rank and select$_1$ in $O(1)$ time using $B(m,n) + O(n)$ bits.*

*Proof.* Assume $n \leq m/2$ – if not, store $S$ in Lemma 1. Let $1 \leq x_1 < \ldots < x_n \leq m$ be the positions of the 1s in the input bit sequence $S$, and equate $S$ with the sequence $\{x_i\}$. Let $b = 2^{\lceil \lg(m/n) \rceil}$, and distribute $S$ into $z = \lceil m/b \rceil = \Theta(n)$ *buckets*, each of length $b$ by placing $x \in S$ into the bucket $\lfloor x/b \rfloor$. We encode the number of elements mapped to each bucket in unary as the bit vector $B = 01^{b_1} 01^{b_2} 0 \ldots 01^{b_z}$, where $b_i$ is the number of 1's in the $i$-th block. We store $B$, whose length is $n + z$, as a FID, and we store an array $L$ where $L[i] = x_i \bmod b$. By Lemma 1, we need $O(n)$ bits to represent $B$, and since $L$ requires $n \lceil \lg b \rceil = n \lg(m/n) + O(n)$ bits, the entire data structure takes $n \lg(m/n) + O(n) = B(m,n) + O(n)$ bits.

The bit vector $B$ demarcates bucket boundaries in $L$, as the $i$-th bucket consists of the elements in $L$ corresponding to the 1s between the indices select$_0(i)$ and select$_0(i+1)$ in $B$. If a bucket has more than $\frac{1}{2c} \frac{\lg m}{\lg \lg m}$ elements in it, then we store a $k$-way search tree on top of the elements of $L$ that belong to the bucket, for $k = \left\lfloor \frac{1}{2c} \frac{\lg m}{\lg \lg m} \right\rfloor$ (the leaves of this search tree are the elements of $L$ and are not stored). Since each bucket has only $O(m/n) = (\lg m)^{O(1)}$ elements in it, and each element in $L$ is of size $\lg(m/n) + O(1) = c \lg \lg m + O(1)$ bits, the search tree is of constant height, and the sum of the sizes of all the search trees over all the buckets is at most $O(n(\lg \lg m)^2/\lg m) = o(n)$ bits.

To support select$_1(i)$, we find the bucket to which the $i$-th 1 belongs, by finding the number of 0s before the $i$-th 1 in $B$ using the FID for $B$. (Specifically, compute select$_1(B,i) - i + 1$.) The remaining lower-order bits are read from the location $L[i]$. To support rank$(i)$, the $i$-th position belongs to the $j$-th block, where $j = \lfloor i/b \rfloor$. We first find the number of 1's upto the beginning of the $j$-th block (by finding the number of 1's upto the $j$-th 0 in $B$ as select$_0(B,j) - j + 1$. We then search for the key $i \bmod b$ in the $j$-th bucket of $L$ using the search tree structure stored (if any). Note that since the elements of $L$ are $O(\lg \lg m)$ bits long, all keys stored at an internal node of the tree can be stored in a single word, and a predecessor search at an internal node can be done in $O(1)$ time using table lookup. If no search tree is stored, then there are no more than $k$ elements in the bucket, and one can perform table lookup using the entire bucket. □

The following lemma is key in what follows. Although it is not substantially new (the key ideas are adapted from [3, 13]) the form of the lemma is particularly convenient for what follows.

**Lemma 3.** *Given integers $N_0, N_1 > 0$ and $M = N_0 + N_1$, such that $\min\{N_0, N_1\} \geq M/(\lg M)^c$ for some constant $c$, we can store a bit vector $S$ with $n_0 \leq N_0$ 0s* and $n_1 \leq N_1$ 1s, using $B(M, N_1) + O(\min\{N_0, N_1\})$ bits, such that $\mathsf{select}_0$ and $\mathsf{select}_1$ are supported in $O(1)$ time.

*Proof.* Assume that $S$ begins with a 1 and ends with a 0 (otherwise adjust Proposition 1 below appropriately). We describe $S$ by two bit-vectors $R_0$ and $R_1$, defined as follows. If there are $r$ runs of 0s of length $l_1, l_2, \ldots, l_r$ in $S$, then $R_0$ is simply $0^{l_1-1}10^{l_2-1}1\ldots0^{l_r-1}1$. $R_1$ is defined analogously, using the runs of 1s (note that there are $r$ runs of 1s as well).

**Proposition 1 ([3]).** *Let $S$ be as above. Then:*

- *$R_0$ and $R_1$ are of length $n_0$ and $n_1$ and both have weight at most $r$.*
- $\mathsf{select}_1(S, i) = \mathsf{select}_1(R0, \mathsf{rank}_1(R1, i-1)) + i$
- $\mathsf{select}_0(S, i) = \mathsf{select}_1(R1, \mathsf{rank}_1(R0, i-1) + 1) + i$,

*taking $\mathsf{select}_1(\cdot, 0) = \mathsf{rank}_1(\cdot, 0) = 0$ on the RHS in the last two bullet points.*

Thus, we only need to support $\mathsf{select}_1$ and $\mathsf{rank}_1$ on $R_0$ and $R_1$. If $N_0 \leq N_1$, then store $R_0$ according to Lemma 1, which takes $n_0 + o(n_0) = O(N_0)$ bits, and store $R_1$ using Lemma 2. We pad out $R_1$ to have exactly $N_1$ 1s and $N_0$ 0s by appending $N_1 - n_1$ 1s and $N_0 - r$ 0s (note that $r \leq N_0$), and represent $R_1$ using Lemma 2 (the padding is done to satisfy the preconditions of the lemma). The space bound is $B(M, N_0) + O(N_0) = B(M, N_1) + O(\min\{N_0, N_1\})$ bits in this case. If $N_1 < N_0$, switch the representations of $R_0$ and $R_1$; the space bound is still the same. □

### 2.2 A succinct index for FID operations on sparse bit vectors

Recall that a systematic encoding of a bit vector $S$ accesses $S$ through $\mathsf{access}$ operations, building an *index* to minimize the calls to $\mathsf{access}$ and to support the operations rapidly. We begin by stating the target we are aiming for – the function $R()$ defined in the next proposition is in fact the same function as that of Theorem 2, with the difference that the lower bound is in the bit-probe model, and the upper bound assumes that each $\mathsf{access}$ operation returns $O(\lg m)$ consecutive bits of $S$. We then use the data structures developed in the previous section to create succinct indices for $\mathsf{rank}$ and $\mathsf{select}_1$, and then for $\mathsf{select}_0$.

**Proposition 2.** *Let $m > 0$ be an integer, and let $n$ be some integer function of $m$, where $1 \leq n < m$. Let $t > 0$ be an integer and let $\mu = \lceil t \lg m \rceil$. Then, if*

$$R(m, n, t) = \begin{cases} \frac{m}{t \lg m} \lg\left(\frac{nt \lg m}{m}\right) & , \text{if } n = \omega(m/(t \lg m)) \\ n\left(1 + \lg\left(\frac{m}{nt \lg m}\right)\right) & , \text{if } n = O(m/(t \lg m)), \end{cases}$$

*then $B(n + \lceil m/\mu \rceil, n) = O(R(m, n, t))$.*

*Proof.* Follows from the standard approximation to the binomial coefficients, namely $\lg \binom{a}{b} = O(b \lg(a/b))$ if $b \leq a/2$.

**A succinct index for rank/select$_1$**

**Lemma 4.** *Given a bit vector $S$ of length $m$ with weight $n$, where $\min\{n, m - n\} \geq m/(\lg m)^c$, for some constant $c > 0$, there is a succinct index that supports* rank *and* select$_1$ *on $S$ that uses $O(t)$ time, $O(t)$* access *calls, and $O(R(m,n,t))$ bits of space, for any $t = (\lg m)^{O(1)}$.*

*Proof.* Partition $S$ into contiguous blocks of size $\mu = \lceil t \lg m \rceil$ each, and let $n_i \geq 0$ denote the number of 1s in the $i$-th block. We represent the sequence $OD = 1^{n_1} 0 1^{n_2} 0 \ldots$, which has $n$ 1s and $\lceil m/\mu \rceil$ 0s using Lemma 3; the index size is $B(n + \lceil m/\mu \rceil, n) + O(\min\{m/\mu, n\}) = O(R(m,n,t))$ bits, as required. Using standard approaches, we can assume that the "local" computation needed to perform rank and select$_1$ operations on a block can be done in $O(t)$ time using pre-computed lookup tables of size $O(m^{2/3})$.

- To compute rank$(S, i)$, let $j = \lfloor i/\mu \rfloor$. If $j = 0$, then the answer is obtained by reading the first block of $S$ with $O(t)$ access operations. Otherwise, select$_0(OD, j) - j$ gives the number of 1s in blocks $1, \ldots, j$, and reading the next block with with $O(t)$ access operations gives the answer to the query.
- To compute select$_1(S, i)$, we first compute $j = $ select$_1(OD, i) - i + 1$, giving us the block in which the $i$-th 1 lies. A call to select$_0(OD, j-1)$ gives the number of 1s in blocks $1, \ldots, j-1$, after which $O(t)$ access calls suffice to compute the answer. □

**A succinct index for select$_0$**

**Lemma 5.** *Given a bit vector $S$ of length $m$ with weight $n$, where $\min\{n, m - n\} \geq m/(\lg m)^c$, for some constant $c > 0$, there is a succinct index that supports* select$_0$ *on $S$ that uses $O(t)$ time, $O(t)$* access *calls, and $O(R(m,n,t))$ bits of space, for any $t = (\lg m)^{O(1)}$.*

*Proof.* We divide $S$ into blocks of size $\mu = \lceil t \lg m \rceil$ as before. Let $x_1 < x_2 < \ldots < x_z$ be the positions of the 0s in $S$ such that rank$_0(S, x_i) = i\mu$, for $i = 1, 2, \ldots, z = \lfloor (m-n)/\mu \rfloor$. Taking $x_0 = 0$, if there are $b_i$ 1s between $x_{i-1}$ and $x_i$, the bit vector $SP = 1^{b_1} 0 1^{b_2} 0 \ldots 1^{b_z} 0$ has at most $n$ 1s and $m/\mu$ 0's, and is represented using at most $B(n + \lceil m/\mu \rceil, n) + O(\min\{n, m/\mu\})$ bits using Lemma 3 so that select$_0$ and select$_1$ are supported in $O(1)$ time. Observe that select$_0(S, i\mu) = i\mu + $ select$_0(SP, i)$, so we now are able to answer select$_0$ queries for the positions $i\mu$, for $i = 1, 2, \ldots, z$.

To answer general select$_0$ queries, we proceed as follows. With each position $x_{i\mu}$ we associate the *gap* $[x_{i\mu}, x_{(i+1)\mu})$. We say that position $x_{i\mu}$ is the starting point of a *long gap* if $x_{(i+1)\mu} - x_{i\mu} + 1 \geq 2\mu$ and define a set $LG$ to be those positions which are the starting point of a long gap (see [2] for a related idea). A key property is that there are at most $n/\mu$ long gaps and that $\sum_{i\mu \in LG} x_{(i+1)\mu} - x_{i\mu} + 1 = O(n)$. This is because any long gap contains at least $\mu$ 1s, and so there are at most $n/\mu$ long gaps; but each long gap always contains $\mu$ 0s, and there are at most $n$ 1s that lie within long gaps.

The bit vector $LG$ (overloading notation), whose $i$-th bit is 1 iff $x_{i\mu}$ is the starting point of a long gap, has $z \leq (m-n)/\mu$ 0s and at most $n/\mu$ 1s. We represent $LG$ using Lemma 2 (if $LG$ has fewer than $n/\mu$ 1s we append 1s to the end, so that $LG$ is not too sparse to apply the lemma). The space used by $LG$ is $B(\lfloor m/\mu \rfloor, \lfloor n/\mu \rfloor) + O(m/\mu)$ bits, which is negligible. Observe that $\mathsf{select}_0(i)$ can be computed in $O(1)$ time if $x_{\mu \cdot \lfloor i/\mu \rfloor}$ is *not* the starting point of a long gap (which can be tested using $LG$), as we can read all the bits in the gap starting at $x_{\mu \cdot \lfloor i/\mu \rfloor}$ using $O(t)$ $\mathsf{access}$ operations, and operate on them in $O(t)$ table-lookups.

We now consider $\mathsf{select}_0$ when the answer is in a long gap. Since there are at most $n/\mu$ long gaps of total length $O(n)$ for some constant $c > 0$, the maximum possible number of blocks $b$ that the long gaps can straddle is at most $O(n/\mu)$ blocks in $S$. Furthermore, the maximum possible number $t$ of 0s in long gaps is $O(n)$. If the $i$th block (partially or fully) contained in a long gap has $z_i$ zeros in it then the bit vector $ZD = 0^{z_1}10^{z_2}1\ldots0^{z_t}1$ is represented using Lemma 3. Observe that $ZD$ has $t$ 0s and $b$ 1s, so its space usage is $O(B(t+b, b)) = O((n/\mu) \lg \mu)$. This is always $O(R(m, n, t))$, since if $n = \omega(m/\mu)$ then $R(m, n, t) = O((m/\mu) \lg \mu)$, and if $n = O(m/\mu)$ then $R(m, n, t) = O(n \lg \mu)$.

The steps to answer $\mathsf{select}_0$ when the answer is in a long gap are as follows:

(a) Let $r = \mu \lfloor i/\mu \rfloor$, and obtain $x_r = \mathsf{select}_0(S, r)$.
(b) If $x_r$ is the starting point of a long gap, then $q = \mathsf{rank}_1(LG, r/\mu)$ gives the the number of long gaps preceding $x_r$.
(c) The number of block boundaries crossed by the interval from $x_r$ to $\mathsf{select}_0(S, i)$ can be obtained by taking the difference in position between 0s corresponding to these in $ZD$ (which is $\mathsf{select}_0(ZD, i \bmod \mu + q\mu) - \mathsf{select}_0(ZD, q\mu)$) and subtracting from it the number of 0s in $ZD$ between these two positions ($i \bmod \mu - 1$). Since we know the block in which $x_r$ lies, we have identifed the block in which $\mathsf{select}_0(S, i)$ lies; it now merely remains to find the number of 0s before this block, which is just a $\mathsf{rank}$ operation and can be answered by using the $OD$ bit vector of Lemma 4, and then reading the block itself in $O(t)$ $\mathsf{access}$ operations and $O(t)$ time. □

Lemmas 4 and 5 show the following main theorem:

**Theorem 1.** *Given a bit vector $S$ of length $m$ with weight $n$, where $\min\{n, m-n\} \geq m/(\lg m)^c$, for some constant $c > 0$, there is a succinct index that supports all FID operations on $S$ that uses $O(t)$ time, $O(t)$ $\mathsf{access}$ calls, and $O(R(m, n, t))$ bits of space, for any $t = (\lg m)^{O(1)}$, where $R(m, n, t)$ is as defined in Proposition 2*

We remark that the condition that $\min\{n, m-n\} \geq m/(\lg m)^c$ is essential to get $O(1)$ time operations, as the predecessor lower bounds of [22] also apply in this setting. Given a set $S \subseteq [m]$, $|S| = n$, we can represent the characteristic vector of $S$ in $O(n)$ words of space in the index by storing the positions of all 1s in the bit vector in the index. Since $R(m, n, t)$ is also at most $O(n)$ words of memory, if we could achieve an index of size $R(m, n, t)$ and support $O(1)$-time $\mathsf{rank}$ operations on the characteristic vector of $S$ for $n = m/(\lg m)^{\omega(1)}$, we would be able to solve predecessor queries in $O(1)$ time for $n = m/(\lg m)^{\omega(1)}$ using $O(n)$ words of memory, which is impossible [22].

## 3 Density-sensitive lower bounds

In this section, we first develop new bounding techniques for binomial coefficients and use them to prove the following theorem.

**Theorem 2.** *The size of the index to support the operations $\mathsf{rank}_1$ or $\mathsf{select}_1$ on bit vectors of length $m$ and weight $n$ satisfies*

$$r = \begin{cases} \Omega\left(\frac{m}{t} \lg\left(\frac{nt}{m}\right)\right) & , \text{if } \frac{nt}{m} = \omega(1) \\ \Omega(n) & , \text{if } \frac{nt}{m} = \Theta(1) \\ \Omega\left(n \lg\left(\frac{m}{nt}\right)\right) & , \text{if } \frac{nt}{m} = o(1). \end{cases}$$

Golynski [10] showed that $r = \Omega((n/t) \lg t)$ for both $\mathsf{rank}_1$ and $\mathsf{select}_1$. This bound is tight only in the case of constant density bit vectors, i.e. when $n = \Theta(m)$. For sparse bit vectors, e.g. when $n < m/t$, the bound of [10] is smaller than optimal by almost a factor of $t$.

In this section, we refine the techniques used in [10] and show tight bounds on the index size for $\mathsf{rank}$ and $\mathsf{select}$ operations in systematic encodings. We prove bounds for the $\mathsf{rank}$ problem, and defer the details of $\mathsf{select}$ to the full version. Consider $\gamma$ queries $\mathcal{Q}^* = \{\text{"}\mathsf{rank}_1(m/\gamma)\text{", "}\mathsf{rank}_1(2m/\gamma)\text{", } \ldots\}$, where $\gamma$ is a parameter which will be chosen later such that $\gamma$ divides $m$. Let $I(B)$ denote the index of size $r$ that is used by the $\mathsf{rank}_1$ algorithm on $B$. Construct the decision tree $T$ for the following procedure: first probe all the $r$ bits stored in $I$, and then simulate the computation of $\mathcal{Q}^*$ queries one by one. The nodes on the first $r$ levels of this tree are labeled by "$I[p] = ?$" for $1 \le p \le r$, and the rest of the nodes are labeled "$B[p] = ?$" for $1 \le p \le m$. The edges are labeled by 0 or 1. Let $x$ be a leaf of $T$. For simplicity of presentation, we perform arbitrary extra probes, so that all the leaves of $T$ are at the same depth $r + t\gamma$. Call $B$ *compatible* with $x$ if $I(B)$ corresponds to the first $r$ edges on the path from the root to $x$, and the probes performed on $B$ by our computation correspond to the rest of the edges on the path. The set of such vectors is denoted by $C(x)$. We note that the bit vectors $B_1, B_2 \in C(x)$ share some common features, e.g. $I(B_1) = I(B_2)$, the locations and the contents of the probed bits by our computation are identical, and the answers to the queries in $\mathcal{Q}^*$ on $B_1$ and $B_2$ are also identical.

The idea of the lower bound proof is as follows. Consider the set $\mathcal{H}$ of $\binom{m}{n}$ bit vectors of length $m$ with $n$ 1-bits. These bit vectors are distributed among the leaves in some fashion. Imagine, that we have a bound $|C(x)| \le C^*(x)$, and let $C^*$ be the sum of $C^*(x)$ across all the leaves. Being an upper bound on the number of leaves, $C^*$ is at least $|\mathcal{H}|$. The bounds derived in [10] are such that $C^* = 2^r D^*$, where $D^*$ does not depend on $r$ (intuitively, $C^*$ is proportional to the number of leaves in $T$). Hence $r$ should be at least $\lg(|\mathcal{H}|/D^*)$.

The bound $C^*(x)$ can be derived as follows. Let us split all the locations in the bit vector into $\gamma$ blocks, the first block spanning positions $1, 2, \ldots, m/\gamma$, the second block spanning positions $m/\gamma+1, m/\gamma+2, \ldots, 2m/\gamma$ and so on. Let $u_i(x)$ be the number of unprobed locations in the $i$-th block in the bit vectors that are compatible with $x$, $y_i(x)$ be the number of 1-probes performed on the block (on the root to leaf path), and $v_i(x) = \mathsf{rank}_1((i+1)m/\gamma) - \mathsf{rank}_1(im/\gamma) - y_i(x)$ be the number of unprobed 1-bits in the block (their locations can be different for different $B \in C(x)$, however the number is fixed for a given leaf, since both $y_i$ and the result of $\mathsf{rank}_1$ queries are known). From now on, we omit parameter $x$ and use just $u_i, v_i, y_i$ to denote these quantities, e.g. define $y := \sum_i y_i$. We have,

$$|C(x)| \leq C^*(x) = \binom{u_1}{v_1}\binom{u_2}{v_2}\cdots\binom{u_\gamma}{v_\gamma}, \tag{1}$$

where $U := \sum u_i = m - t\gamma$ (since exactly $t\gamma$ positions are probed for each leaf) and $V := \sum v_i = n - y$ (since $y$ is the total number of probed 1-bits). Let $L_y$ be the group of leaves for which there are exactly $y$ 1-probes. Note that $|L_y| = 2^r \binom{t\gamma}{y}$. For each $y$, let $x_y$ be the leaf in $L_y$ that maximizes the product (1). Hence, partitioning all $2^{r+t\gamma}$ leaves w.r.t. $y$, we have,

$$C^* \leq 2^r \sum_{y=0}^{\min\{t\gamma, n\}} \binom{t\gamma}{y} C^*(x_y) \leq n 2^r X,$$

where $X$ is the maximum of $\binom{t\gamma}{y}\binom{u_1}{v_1}\binom{u_2}{v_2}\cdots\binom{u_\gamma}{v_\gamma}$ over all possible choices of $y$, $u_i$'s and $v_i$'s, such that $t\gamma + \sum_i u_i = m$, $y + \sum_i v_i = n$, $0 \leq u_i \leq m/\gamma$, and $0 \leq v_i \leq u_i$. The bounding methods of [10] are too crude for our purposes, so we first need to develop better bounding techniques.

**Lemma 6 (Stirling's approximation).** *For $n \geq 1$, we have*

$$\sqrt{2\pi} < \frac{n!}{\sqrt{n}(n/e)^n} \leq e.$$

**Lemma 7.** *For values $u$ and $v$, such that $0 < v \leq u/2$, we have*

$$\frac{1}{e} < \frac{\binom{u}{v}}{\frac{1}{\sqrt{v}}\left(\frac{u}{v}\right)^v \left(\frac{u}{u-v}\right)^{u-v}} < \frac{4}{5}.$$

*Proof.* We start by estimating the value of $u!/(u-v)!$. To do so, we first show that the sequence $a_n = n!/(n/e)^n$ is increasing and $b_n = n!/(n/e)^{n+1}$ is decreasing for integers $n$, $n > 0$. Consider

$$\frac{a_{n+1}}{a_n} = \frac{(n+1)!e^{n+1}}{(n+1)^{n+1}} \frac{n^n}{n!e^n} = \frac{e}{\left(1+\frac{1}{n}\right)^n} = e^{1-n\lg(1+1/n)} > 1,$$

since $\lg(1+x) < x$ for $x > -1$. In a similar fashion, consider

$$\frac{b_{n+1}}{b_n} = \frac{(n+1)!e^{n+2}}{(n+1)^{n+2}} \frac{n^{n+1}}{n!e^{n+1}} = e\left(1 - \frac{1}{n+1}\right)^{n+1} = e^{1+(n+1)\lg(1-1/(n+1))} < 1,$$

since $\lg(1-x) < -x$ for $x < 1$. Since $a_u/a_{u-v} > 1$ and $b_u/b_{u-v} < 1$, we have

$$\frac{u!}{(u-v)!} = \frac{\frac{a_u u^u}{e^u}}{\frac{a_{u-v}(u-v)^{u-v}}{e^{u-v}}} > \frac{(u/e)^u}{((u-v)/e)^{u-v}} = \left(\frac{u}{e}\right)^v \left(\frac{u}{u-v}\right)^{u-v}, \text{ and}$$

$$\frac{u!}{(u-v)!} = \frac{\frac{b_u u^{u+1}}{e^{u+1}}}{\frac{b_{u-v}(u-v)^{u-v+1}}{e^{u-v+1}}} < \frac{u}{u-v}\frac{(u/e)^u}{((u-v)/e)^{u-v}} = \frac{u}{u-v}\left(\frac{u}{e}\right)^v \left(\frac{u}{u-v}\right)^{u-v}.$$

We divide both of these inequalities by $v!$, and use Lemma 6. We obtain

$$\frac{1}{e}\frac{1}{\sqrt{v}}\left(\frac{e}{v}\right)^v\left(\frac{u}{e}\right)^v\left(\frac{u}{u-v}\right)^{u-v} < \binom{u}{v} < \frac{1}{\sqrt{2\pi}}\frac{1}{\sqrt{v}}\frac{u}{u-v}\left(\frac{e}{v}\right)^v\left(\frac{u}{e}\right)^v\left(\frac{u}{u-v}\right)^{u-v}$$

By the precondition of the lemma, $v \leq u/2$, so that $u/(u-v) \leq 2$. Also, since $\sqrt{2\pi} > 5/2$, the statement of the lemma follows. □

Let us define $u_* = \min u_i$ and $v_* = \min v_i$. In the case where $v_i$'s are of the same order of magnitude, we can use the following lemma.

**Lemma 8.** *If $u_* \geq 2$ and $v_* \geq 1$, then $\prod_i \binom{u_i}{v_i} \leq \binom{U}{V} 2^{-(\gamma/2)\lg v_* - 0.3\gamma + (\lg V)/2}$.*

*Proof.* To bound each individual multiplier, we apply the right part of the inequality of Lemma 7. If $v_i \leq u_i/2$, then

$$\prod_i \binom{u_i}{v_i} \leq \prod_i \left(\frac{4}{5\sqrt{v_i}}\right)\left(\frac{u_i}{v_i}\right)^{v_i}\left(\frac{u_i}{u_i - v_i}\right)^{u_i - v_i}.$$

The case where $v_i > u_i/2$ can be reduced to the case $v_i = u_i/2$. Next, we apply the inequality between arithmetic and geometric means for the values

$$\underbrace{\frac{u_1}{v_1}, \ldots, \frac{u_1}{v_1}}_{v_1 \text{ times}}, \underbrace{\frac{u_2}{v_2}, \ldots, \frac{u_2}{v_2}}_{v_2 \text{ times}}, \ldots \underbrace{\frac{u_\gamma}{v_\gamma}, \ldots, \frac{u_\gamma}{v_\gamma}}_{v_\gamma \text{ times}}, \text{ and obtain}$$

$$\prod_i \left(\frac{u_i}{v_i}\right)^{v_i} \leq \left(\frac{\sum_i v_i \cdot \frac{u_i}{v_i}}{\sum_i v_i}\right)^{\sum_i v_i} = \left(\frac{U}{V}\right)^V. \text{ Similarly, we obtain}$$

$$\prod_i \left(\frac{u_i}{u_i - v_i}\right)^{u_i - v_i} \leq \left(\frac{\sum_i u_i}{\sum_i u_i - v_i}\right)^{\sum_i u_i - v_i} = \left(\frac{U}{U-V}\right)^{U-V}.$$

Finally, we apply the left part of the inequality of Lemma 7,

$$\prod_i \binom{u_i}{v_i} < e\sqrt{V}\binom{U}{V}\prod_i \frac{4}{5\sqrt{v_i}} \leq 2^{-(\gamma/2)\lg v_* - 0.3\gamma + (\lg V)/2}\binom{U}{V}.$$

□

To bound the product $f(v_1, v_2, \ldots, v_\gamma) := \prod_i \binom{u_i}{v_i}$, we maximize it over all possible $v_i$'s with $u_i$'s fixed, subject to the constraint that the sum of $v_i$'s is $V$. We say that a tuple $(v_1, v_2, \ldots, v_\gamma)$ is a *local maximum* if we can not increase the value of $f$ by changing some $v_i$ to $v_i + 1$ and some other $v_j$ to $v_j - 1$. The following lemma characterizes the local maxima.

**Lemma 9.** *At a local maximum,*

$$\frac{v_j + 1}{u_j + 1} \geq \frac{v_i}{u_i + 1}$$

*is satisfied for each pair $(i, j)$, $i \neq j$.*

*Proof.* At a local maximum, we have the following inequality

$$\binom{u_i}{v_i}\binom{u_j}{v_j} \geq \binom{u_i}{v_i - 1}\binom{u_j}{v_j + 1}$$

Dividing both parts by $\frac{(u_i)!}{(v_i-1)!(u_i-v_i)!} \frac{(u_j)!}{(v_j)!(u_j-v_j-1)!}$, we obtain

$$\frac{1}{v_i} \frac{1}{(u_j - v_j)} \geq \frac{1}{(u_i - v_i + 1)} \frac{1}{(v_j + 1)}$$

From this, we get $u_i v_j + u_i - v_i + v_j + 1 \geq v_i u_j$ and the lemma follows. □

**Corollary 1.** *At a local maximum, for all $i$, $1 \leq i \leq \gamma$, we have*

$$\left| \frac{v_i}{u_i} - \frac{V}{U} \right| < \frac{2}{u_*}.$$

*Proof.* Fix $i \neq j$, and apply Lemma 9 for the pair $(i, j)$ and for the pair $(j, i)$. It follows that

$$\frac{v_j}{u_j + 1} + \frac{1}{u_j + 1} \geq \frac{v_i}{u_i + 1} \geq \frac{v_j}{u_j + 1} - \frac{1}{u_i + 1}$$

Since $u_i$ and $u_j$ are at least $u_*$, we have

$$\left| \frac{v_i}{u_i + 1} - \frac{v_j}{u_j + 1} \right| \leq \frac{1}{u_* + 1} < \frac{1}{u_*}$$

Since $v_i/u_i$ and $v_j/u_j$ are at most 1,

$$\left| \left( \frac{v_i}{u_i + 1} - \frac{v_j}{u_j + 1} \right) - \left( \frac{v_i}{u_i} - \frac{v_j}{u_j} \right) \right| = \left| \frac{v_i/u_i}{u_i + 1} - \frac{v_j/u_j}{u_j + 1} \right| < \frac{1}{u_*}$$

Therefore,

$$\left| \frac{v_i}{u_i} - \frac{v_j}{u_j} \right| < \frac{2}{u_*}.$$

Finally, we observe that

$$\min \left\{ \frac{v_1}{u_1}, \ldots, \frac{v_\gamma}{u_\gamma} \right\} \leq \frac{V}{U} \leq \max \left\{ \frac{v_1}{u_1}, \ldots, \frac{v_\gamma}{u_\gamma} \right\}$$

and the corollary follows. □

**Lemma 10.** *If $u_*V/U \geq 3$, then $\prod_i \binom{u_i}{v_i} \leq \binom{U}{V} 2^{-(\gamma/2)\lg(u_*V/U) - 0.3\gamma + (\lg V)/2}$.*

*Proof.* We first maximize $X$ with respect to $v_i$'s for fixed $u_i$'s. At a local maximum, Corollary 1 gives us the bound

$$\frac{v_i}{u_i} > \frac{V}{U} - \frac{2}{u_*} > \frac{V}{U} - \frac{2V}{3U} = \frac{V}{3U},$$

so that $v_i > u_i V/(3U)$. Hence, we can apply Lemma 8 with $v_* = u_i V/(3U) \geq u_* V(3U) \geq 1$. The result follows. □

### 3.1 Density-Sensitive Rank Index

(Theorem 2 for the rank$_1$ operation).

*Proof.* Let us define $k := m/\gamma$ to be the length of a block. We combine consecutive blocks into larger *superblocks*, such that the number of unprobed bits in the $i$-th superblock, $u_i^*$, is between $k$ and $2k$ (except, possibly, for the last superblock). This can be done in a greedy fashion, considering blocks from left to right: we keep adding blocks to a superblock until the number of unprobed bits in it reaches $k$, at which point we finalize it and start a new one. We will never overshoot the value $2k$, since all $u_i$'s are at most $k$. It was shown in [10] that the number of superblocks $\gamma_s = \Theta(\gamma)$, and $\prod_i \binom{u_i}{v_i} \leq \prod_i \binom{u_i^*}{v_i^*}$, where $v_i^*$ is the number of unprobed 1-bits in the $i$-th superblock.

First, consider the case $tn \geq m$. Let us choose $\gamma$ to be $m/(3t)$. We can apply Lemma 10 to $\binom{t\gamma}{y} \prod_i \binom{u_i^*}{v_i^*}$, since $n \min\{t\gamma, \min\{u_i^*\}\}/m = \min\{n/3, kn/m\} = \min\{n/3, 3tn/m\} \geq 3$. Recalling that $U = m - t\gamma$ and $V = n - y$, so that $\binom{t\gamma}{y}\binom{U}{V} < \binom{m}{n}$, we obtain

$$C^* \leq n2^r \binom{t\gamma}{y} \prod_{i=1}^{\gamma_s} \binom{u_i^*}{v_i^*} \leq n2^r 2^{-(\gamma_s/2)\lg(3tn/m) - 0.3\gamma_s + (\lg n)/2} \binom{m}{n}. \text{ Hence,}$$

$$r \geq (\gamma_s/2)\lg(3tn/m) + 0.3\gamma_s - 3(\lg n)/2 = \Omega((m/t)\lg((nt)/m)).$$

If $cm < tn < m$ for some positive constant $c$, then pick $\gamma = n/3$. We have, $n\min\{t\gamma, k\}/m \geq \min\{cn/3, 3\} \geq 3$, and obtain

$$C^* \leq n2^r \binom{t\gamma}{y} \prod_{i=1}^{\gamma_s} \binom{u_i^*}{v_i^*} \leq n2^r 2^{-(\gamma_s/2)\lg 3 - 0.3\gamma_s + (\lg n)/2} \binom{m}{n}, \text{ and}$$

$$r \geq (\gamma_s/2)\lg 3 + 0.3\gamma_s - 3(\lg n)/2 = \Omega(n).$$

Finally, if $nt = o(m)$, then we pick $\gamma = \sqrt{nm/t}$.

We bound the product $\prod_i \binom{u_i}{v_i} \leq \prod_i k^{v_i} \leq (m/\gamma)^V \leq \binom{mV/\gamma}{V} \leq \binom{mn/\gamma}{n-y}$ directly using simple inequalities $(u/v)^v \leq \binom{u}{v} \leq u^v$.

Thus,

$$C^* \leq 2^r \sum_{y=0}^{\min\{t\gamma,n\}} \binom{t\gamma}{y} \prod_i \binom{u_i}{v_i} \leq 2^r \sum_y \binom{t\gamma}{y}\binom{\frac{nm}{\gamma}}{n-y} = 2^r \binom{t\gamma + \frac{nm}{\gamma}}{n} \leq 2^r \binom{\sqrt{nmt}}{n},$$

since $\sqrt{nmt} = \omega(nt) > 2n$.

We can bound $r$ by

$$r \geq \lg\left(\frac{\binom{n}{m}}{\binom{\sqrt{nmt}}{m}}\right) \geq \lg\left(\frac{\left(\frac{n}{m}\right)^m}{\left(\frac{\sqrt{nmt}}{m}\right)^m}\right) \geq m\lg\sqrt{\frac{n}{mt}} = \frac{m}{2}\lg\left(\frac{n}{mt}\right).$$

□

### 3.2 Density-Sensitive Select Index

(Theorem 2 for the $\mathsf{select}_1$ operation).

*Proof.* As with the proof for the case of $\mathsf{rank}_1$ operation, we consider three cases: where $nt = \omega(m)$, $nt = \Theta(m)$, and where $nt = o(m)$. We simulate the set of queries

$$\mathcal{Q}^* = \{\text{"query } \mathsf{select}_{1B}(1, ik)\text{"} | \, 1 \leq i \leq \gamma\},$$

where $k = \lfloor n/\gamma \rfloor$. Accordingly, we split bit vectors $B$ into $\gamma$ *blocks* of equal cardinalities $n_1 = n_2 = \ldots = n_\gamma = k$. The $i$-th block starts at position $\mathsf{select}_1((i-1)k)+1$ and ends at position $\mathsf{select}_1(ik)$, so that the *cardinality* of each block (the number of 1-bits in it) is exactly $k$ (recall that we defined $\mathsf{select}_1(0) = 0$ for convenience). We set $\mathcal{H} = \{B \in \{0,1\}^m | \text{ number of 1-bits in } B \text{ is } n\}$. We choose parameter $\gamma$ depending on the relationship between $nt$ and $m$. In the case where $nt = \omega(m)$, we will choose $\gamma = m/(3t)$. For the case $mt = \Theta(n)$, we need an additional requirement that $\gamma \leq n/3$, so that we will choose $\gamma = \min\{m/(3t), n/3\}$, we will clarify this requirement later in the proof. Finally, for the case $nt = o(m)$, we will choose $\gamma = n$. Note that in all cases, the number of unprobed bits $U$ is $m - t\gamma \geq 2m/3$, so that the average number of unprobed bits per block is at least $(2/3)n/\gamma$ (we expect most of the blocks to have at least constant fraction of unprobed bits).

We define *superblocks* as follows. The $i$-th superblock (except, perhaps, for the last one) will contain consecutive blocks $z_{i-1} + 1, \ldots, z_i$, such that the number of unprobed 1-bits in the $i$-th superblock

$$v_i^* = v_{z_{i-1}+1} + v_{z_{i-1}+2} + \ldots + v_{z_i}$$

satisfies $k \leq v_i^* < 2k$. Note that this is always possible, since $v_i \leq n_i = k$. And hence, $\gamma_s$, the number of superblocks, is at least $V/(2k)$. The number of unprobed bits of $i$-th superblock is given by

$$u_i^* = u_{z_{i-1}+1} + u_{z_{i-1}+2} + \ldots + u_{z_i}$$

We use inequality

$$\binom{u_1}{v_1}\binom{u_2}{v_2}\cdots\binom{u_\gamma}{v_\gamma} \leq \binom{u_1^*}{v_1^*}\binom{u_2^*}{v_2^*}\cdots\binom{u_{\gamma_s}^*}{v_{\gamma_s}^*}$$

to bound the number of bit vectors compatible with a given leaf. The total number of bit vectors compatible with all the leaves is then

$$P := 2^r \sum_{y=0}^{\min\{t\gamma,n\}} \binom{t\gamma}{y} \binom{u_1^*}{v_1^*} \binom{u_2^*}{v_2^*} \cdots \binom{u_{\gamma_s}^*}{v_{\gamma_s}^*} \quad (2)$$

We can derive a bound on $P$, $P \leq n 2^r X$, where $X$ is the biggest product of binomial coefficients in this sum. To derive a bound on $X$, we can, for example, use Lemma 8. A difference with the proof of the $\mathsf{rank}_1$ case is that we do not need to "redistribute" the weight of $V$ between $v_i$'s uniformly as it was done in Lemma 9 and Corollary 1, since we have bounds $k \leq v_i^* < 2k$ already. To derive a lower bound for $r$, we observe that $\sum_x |C(x)| = |\mathcal{H}| = \binom{m}{n} \leq P$. Therefore,

$$r \geq \lg\left(\frac{\binom{m}{n}}{X}\right) - \lg n.$$

- First, we consider the case where $nt/m = \Omega(1)$. Recall that we chose the parameter $\gamma = \min\{m/(3t), n/3\} = \Theta(m/t)$. The goal is to derive an upper bound on

$$X = \binom{t\gamma}{y} \binom{u_1^*}{v_1^*} \binom{u_2^*}{v_2^*} \cdots \binom{u_{\gamma_s}^*}{v_{\gamma_s}^*} \quad (3)$$

subject to constraints

$$t\gamma + \sum_i u_i^* = m,$$

$$y + \sum_i v_i^* = n, \text{ and}$$

$$t\gamma \leq \frac{m}{3}$$

Since there is a bound on $v_i^*$'s, namely, $v_i^* \geq k$, it seems that we can apply Lemma 8 directly and obtain a bound on $X$. The caveat is that, if $V$ is too small, then the number of superblocks $\gamma_s$ is small as well, and the bound will turn out to be weak. This problem did not arise in the proof of the $\mathsf{rank}_1$ case, since the bound on $\gamma_s$ was based on the fact that $\sum u_i \geq 2n/3$, and we were grouping blocks into superblocks based on values of $u_i$. However, in this proof, we form superblocks based on $v_i$'s, so that we need to bound their sum, $V = \sum_i v_i$, from below.

For this purpose, we use the idea that is similar to an idea in the proof of Lemma 10. Let us vary $u_i$'s and $v_i$'s in order to maximize

$$\binom{t\gamma}{y} \prod_i \binom{u_i}{v_i}.$$

As a very rough estimation, we can state the following: since $t\gamma \leq m/3$, we expect that $y$ will be at most $n/3$, and so that $V = n - y \geq 2n/3$, which is sufficient for our purposes. More formally, Lemma 9 gives us the following conditions at a local maximum:

$$\frac{v_i + 1}{u_i + 1} \geq \frac{y}{t\gamma + 1} \geq \frac{y}{m/3 + 1}$$

Thus, for any $i \in [\gamma]$, we have

$$\left(\frac{m}{3} + 1\right)(v_i + 1) \geq y(u_i + 1).$$

Summing them up, we obtain

$$V \geq y\frac{U + \gamma}{m/3 + 1} - \gamma \geq (n - V)\frac{2m/3 + 2}{m/3 + 1} - \frac{n}{3} = \frac{5n}{3} - 2V,$$

since $2 \leq \gamma \leq n/3$. Thus,

$$V \geq \frac{5n}{9}.$$

Now, it is easy to derive a bound on the number of superblocks, $\gamma_s$,

$$\gamma_s \geq \frac{V}{2k} \geq \frac{5n}{18}\frac{\gamma}{n} \geq \frac{5\gamma}{18} = \Theta\left(\frac{m}{t}\right).$$

Let us apply Lemma 8 to (3). We obtain,

$$X \leq \binom{t\gamma}{n}2^{-(\gamma_s/2)\lg k - \Theta(\gamma_s)}\binom{U}{V} \leq 2^{-(\gamma_s/2)\lg k - \Theta(\gamma_s)}\binom{m}{n}$$

$$\leq 2^{-\Theta(m/t)\lg k - \Theta(m/t) + \Theta(\lg n)}\binom{m}{n}.$$

So that, in the case where $nt = \omega(m)$, we obtain

$$r = \Theta\left(\frac{m}{t}\lg\left(\frac{nt}{m}\right)\right) - \Theta\left(\frac{m}{t}\right),$$

since $k = n/\gamma = 3nt/m$. In the case where $nt = \Theta(m)$, we obtain

$$r = \Theta(n),$$

since $k = n/\gamma \geq 3$, $k = \Theta(1)$.

– It remains to consider the case where $mt = o(n)$. Recall that we chose $\gamma = n$ in this case, so that we select all the 1-bits in the bit vector using our queries. Hence, the number of compatible bit vectors $|C(x)|$ with any leaf $x$ is exactly 1. We can bound the sum

$$\sum_x |C(x)| \leq 2^r \sum_{y=0}^{n}\binom{t\gamma}{y} \leq n2^r\binom{tn}{n} \leq n2^r\left(\frac{etn}{n}\right)^n = n2^r(et)^n$$

$$\leq n2^r(et)^n\frac{\binom{m}{n}}{\left(\frac{m}{n}\right)^n} \leq 2^r 2^{-n\lg(m/(nt)) + \Theta(n)}\binom{m}{n},$$

here we used the fact that $\binom{u}{v} \leq (eu/v)^v$. Thus,

$$r = \Theta\left(n \lg \frac{m}{nt}\right) - \Theta(n)$$

This completes the proof of Theorem 2 for select$_1$ operation. □

## 4 Conclusions

We have provided matching density-sensitive upper and lower bounds on the redundancy required for rank and select operations on sparse bit vectors in the succinct index model. These results improve signficantly on known results for the case where $n = o(m)$. Although our results depend on both $m$ and $n$, we do not take into account any regularities in the distribution of 1s in the bit vector; an interesting direction of research would be to find appropriate measures of the regularities of 1s in the bit vector that could lead to further reductions in the index size.